\documentclass[%
 reprint,
 amsmath,amssymb,
 aps,
]{revtex4-2}

\usepackage{graphicx}
\usepackage{subcaption}
\usepackage{overpic}     
\usepackage{dcolumn}
\usepackage{bm}
\usepackage{hyperref}
\usepackage{xcolor}  

\begin{document}

\preprint{APS/123-QED}

\title{First-principles prediction of high-temperature superconductivity in stretched carbon nanotubes}

\author{Hua-Zhen Li$^{1,2}$}
\author{Xun-Wang Yan$^{3}$}\thanks{Corresponding author: \href{mailto:yanxunwang@163.com}{yanxunwang@163.com}}

\affiliation{$^{1}$School of Physics, Renmin University of China, Beijing 100872, China}
\affiliation{$^{3}$School of Physics and Engineering, Qufu Normal University, Qufu 273165, China}

\date{\today}

\begin{abstract}
	Superconductivity in quasi-one-dimensional systems is an significant but undervalued research field.
In this work, based on the electron-phonon coupling mechanism, we systematically investigate the superconductivity in quasi-one-dimensional carbon nanotube under uniaxial tensile strain. The calculated superconducting critical temperature attains its peak value of 162 K at a uniaxial tensile strain of 4.5\%, being drastically higher than the counterpart in the unstrained carbon nanotube.
An overall softening of phonons, strong electron-phonon coupling, and an increase of electronic density of states at the Fermi level, play key roles in achieving high-temperature superconductivity in this system.
Our research demonstrates that stretching is an effective approach to modulating the superconductivity one-dimensional materials, and more importantly, indicates that high-temperature superconductivity may occur in carbon nanotubes.
\end{abstract}

\maketitle


\section{\label{sec:level1}Introduction}
The superconductivity in quasi-one-dimensional materials remains far less explored. This is partly because most superconducting research has long focused on bulk materials, while the research on superconductivity in two-dimensional materials has just emerged over the past few years.
On the other hand, for the quasi-one-dimensional materials, their strongly anisotropic crystal structure, anisotropic electronic structure, and ease of manipulation arising from sensitivity to lattice distortions make such systems an excellent platform for investigating superconductivity.

Carbon-based materials stand as one of the vital members in the family of superconducting materials.
Carbon exhibits a wealth of allotropic structures in nature, which can be classified by dimensionality into zero-dimensional (0D) fullerenes, one-dimensional (1D) carbon nanotubes, two-dimensional (2D) graphene, three-dimensional (3D) diamond, and graphite intercalation structures.
What is striking is that superconductivity has been observed in all these distinct structures. Examples include twisted graphene bilayer \cite{cao2018unconventional}, fullerides \cite{gunnarsson1997superconductivity,takabayashi2009disorder}, graphite intercalation compounds such as CaC$_6$ \cite{emery2005superconductivity}.
Recent theoretical advances have revealed that extended carbon-cage networks can host strong electron-phonon coupling(EPC), with C$_{18}$, C$_{24}$, and C$_{32}$ cage networks exhibiting superconductivity above 100~K \cite{hai2023superconductivity, ye2025superconducting}.
Moreover, potassium-intercalated picene (K$_x$picene) and phenanthrene (K$_x$phenanthrene) shows superconductivity similar to the Ca-intercalated graphite and K-intercalated fullerenes \cite{mitsuhashi2010superconductivity, wang2011superconductivity}, which are members of carbon-based materials, known as aromatic organic superconductors. Among these carbon-based materials, carbon nanotubes stand out for their quasi-one-dimensional structure, which endows them with highly tunable electronic and structural properties.

Research on superconductivity in carbon nanotubes can be traced back to as early as 2001, when Tang~ et al. observed a superconducting transition around 15 K in carbon nanotubes confined within the channels of AlPO$_4$-5 \cite{tang2001superconductivity}. In the same year, Kociak~ et al.  detected a resistive anomaly near 0.55 K in carbon nanotube ropes \cite{kociak2001superconductivity}. More strikingly, a very recent experimental study reported superconductivity in boron-doped networks of ultrathin carbon nanotubes under ambient pressure, with the critical temperature \(T_c\) ranging from 220 to 250 K \cite{wang2025high}.
In terms of theoretical research, Bohnen et al. investigated the superconductivity of the (3,3) carbon nanotube via density functional theory (DFT) calculations in 2004, and the superconducting critical temperature \(T_c\) was found to be 3 K \cite{bohnen2004lattice}.
Meanwhile, Meng et al. conducted high-throughput first-principles calculations of element-doped (3,3) carbon nanotubes, finding that the undoped (3,3) carbon nanotube lacks superconductivity, but Si-doping can induce superconductivity with T$_c$ up to ~28 K \cite{bo2021high}. Very recently, Ouyang \textit{et al.} showed that pristine ultrathin (3,0) carbon nanotubes can support strong electron–phonon coupling (EPC) and phonon-mediated superconductivity with $T_c$ up to $\sim 33$~K under ambient pressure, based on first-principles calculations \cite{ouyang2025electron}.

Strain engineering provides an experimentally accessible route for tuning the electronic and vibrational properties of low-dimensional materials. Furthermore carbon nanotubes are particularly well suited for such strain-based modulation: their exceptional mechanical robustness allows them to sustain several percent of strain without fracture \cite{yakobson1996nanomechanics}, and their quasi-one-dimensional electronic structure makes them highly sensitive to lattice deformation.
Uniaxial strain effectively can be regarded as a form of ``negative pressure''
, reducing the effective lattice density and shifting electronic and vibrational energies in a manner opposite to hydrostatic compression.\cite{yang1999band}.
This is exactly the opposite approach compared with the high-pressure research on hydride superconductivity.
Previous studies have shown that strain can shift van Hove singularities, reconstruct energy bands, soften selected phonon branches, and modify EPC in carbon-based nanostructures \cite{yang2000electronic}.

We focus on the study of superconductivity in the (3,3) carbon nanotube with different uniaxial tensile strain based on the first-principles calculation method.
			
\section{\label{sec:level1}Computational Methods}

All structural models of the (3,3) carbon nanotube were constructed and visually inspected using the VESTA package \cite{vesta}. First-principles calculations were carried out within DFT \cite{dft} using the generalized gradient approximation of Perdew, Burke, and Ernzerhof (PBE) \cite{pbe}, as implemented in the Vienna \textit{Ab initio} Simulation Package (VASP) \cite{vasp1,vasp2}. The projector augmented-wave (PAW) method \cite{paw} was employed to describe electron–ion interactions. A plane-wave cutoff energy of 600~eV and a $1\times1\times256$ Monkhorst-Pack $k$-mesh were used for Brillouin-zone sampling. Electronic self-consistency was achieved with an energy tolerance of $10^{-6}$~eV, and atomic positions were relaxed until residual forces were below $10^{-3}$~eV/\AA.

Vibrational and EPC properties were computed using density functional perturbation theory (DFPT) \cite{baroni2001} as implemented in \textsc{Quantum~Espresso} \cite{qe}. Ultrasoft pseudopotentials from the SG15 Optimized Norm-Conserving Vanderbilt library \cite{ONCV,sg15_ONCV} were used for carbon, incorporating the $2s^2 2p^2$ valence configuration. Wavefunction and charge-density cutoff energies of 100~Ry and 800~Ry were adopted, respectively. Self-consistent electronic structure calculations were performed using a series of $1\times1\times N_k$ Monkhorst--Pack meshes along the nanotube axis, with $N_k$ ranging from 32 to 192, combined with Methfessel--Paxton smearing, in order to assess the convergence of electronic and electron--phonon coupling properties. While dynamical matrices were obtained on a $1\times1\times4$ $q$-point grid. Electronic and force convergence thresholds were set to $10^{-9}$~Ry and $10^{-6}$~Ry/Bohr. The Eliashberg spectral function $\alpha^2F(\omega)$ and the EPC parameter $\lambda$ were evaluated from Brillouin-zone integrations of the DFPT matrix elements.

The electron-phonon matrix element, which governs the scattering amplitude between electronic states $| \psi_{i,\mathbf{k}} \rangle$ and $| \psi_{j,\mathbf{k+q}} \rangle$ due to a phonon of branch $\nu$ and wave vector $\mathbf{q}$, is given by \cite{Allen19722577}
\begin{equation}
	g^{ij}_{\mathbf{k},\mathbf{q}\nu}
	=
	\left(
	\frac{\hbar}{2M\omega_{\mathbf{q}\nu}}
	\right)^{\frac{1}{2}}
	\left\langle
	\psi_{i,\mathbf{k}}
	\left|
	\frac{d V_{\mathrm{SCF}}}{d \hat{u}_{\mathbf{q}\nu}}
	\cdot \hat{e}_{\mathbf{q}\nu}
	\right|
	\psi_{j,\mathbf{k+q}}
	\right\rangle ,
\end{equation}
where $M$ is the ionic mass, $\omega_{\mathbf{q}\nu}$ the phonon frequency, $\hat{e}_{\mathbf{q}\nu}$ the polarization vector, and $V_{\mathrm{SCF}}$ the self-consistent potential.

The phonon linewidth associated with EPC is expressed as
\begin{equation}
	\gamma_{\mathbf{q}\nu}
	=
	\frac{2\pi \omega_{\mathbf{q}\nu}}{\Omega_{\mathrm{BZ}}}
	\sum_{ij}
	\int d^3 k\,
	\left|
	g^{ij}_{\mathbf{k},\mathbf{q}\nu}
	\right|^2
	\delta(\epsilon_{\mathbf{q},i}-\epsilon_F)
	\delta(\epsilon_{\mathbf{k+q},j}-\epsilon_F) ,
\end{equation}
with $\epsilon_F$ the Fermi energy and $\Omega_{\mathrm{BZ}}$ the Brillouin-zone volume.

The Eliashberg spectral function is defined as \cite{RevModPhys.89.015003}
\begin{equation}
	\alpha^2F(\omega) =
	\frac{1}{2\pi N(\epsilon_F)}
	\sum_{\mathbf{q}\nu}
	\delta( \omega - \omega_{\mathbf{q}\nu} )
	\frac{\gamma_{\mathbf{q}\nu}}{\hbar \omega_{\mathbf{q}\nu}},
\end{equation}
where $N(\epsilon_F)$ is the electronic density of states at the Fermi level($E_F$). In numerical evaluations, the electronic $\delta$-functions were replaced with Gaussian smearing of 0.02~Ry. The EPC constant is obtained through
\begin{equation}
	\lambda = 2 \int_0^\infty \frac{\alpha^2F(\omega)}{\omega} d\omega .
\end{equation}

The superconducting critical temperature was estimated using the Allen–Dynes modified McMillan formula \cite{PhysRevB.12.905},
\begin{equation}
	T_c = f_1 f_2 \frac{\omega_{\log}}{1.2}
	\exp \!\left[
	-\frac{1.04(1+\lambda)}{\lambda - \mu^*(1+0.62\lambda)}
	\right],
\end{equation}
where the Coulomb pseudopotential was fixed at $\mu^* = 0.10$. The correction factors are
\begin{equation}
	f_1 = \left[ 1 + \left( \frac{\lambda}{2.46(1+3.8\mu^*)} \right)^{3/2} \right]^{1/3},
\end{equation}
\begin{equation}
	f_2 = 1 +
	\frac{
		\left( \frac{\bar{\omega}_2}{\omega_{\log}} - 1 \right) \lambda^2
	}{
		\lambda^2 +
		3.312(1+6.3\mu^*)^2\left( \frac{\bar{\omega}_2}{\omega_{\log}} \right)^2
	}.
\end{equation}

The logarithmic and second-moment phonon frequencies are defined as
\begin{equation}
	\omega_{\log} =
	\exp\!\left[
	\frac{2}{\lambda}
	\int_0^\infty
	\frac{\alpha^2F(\omega)}{\omega}
	\ln(\omega)\, d\omega
	\right],
\end{equation}
\begin{equation}
	\bar{\omega}_2 =
	\left[
	\frac{2}{\lambda}
	\int_0^\infty
	\omega \alpha^2F(\omega)\, d\omega
	\right]^{1/2}.
\end{equation}

For comparison, the original McMillan formula is recovered by setting $f_1 f_2 = 1$:
\begin{equation}
	T_c = \frac{\omega_{\log}}{1.2}
	\exp\!\left[
	-\frac{1.04(1+\lambda)}{\lambda - \mu^*(1+0.62\lambda)}
	\right].
\end{equation}

\section{\label{sec:level1}Results and Discussion }	
\subsection{\label{sec:level2}Crystal structure and structural stability under the unstrained state}

\begin{figure}[t]
	\centering
	\begin{minipage}{0.85\columnwidth}
			\begin{overpic}[trim=18 0 5 0, clip, width=\textwidth]{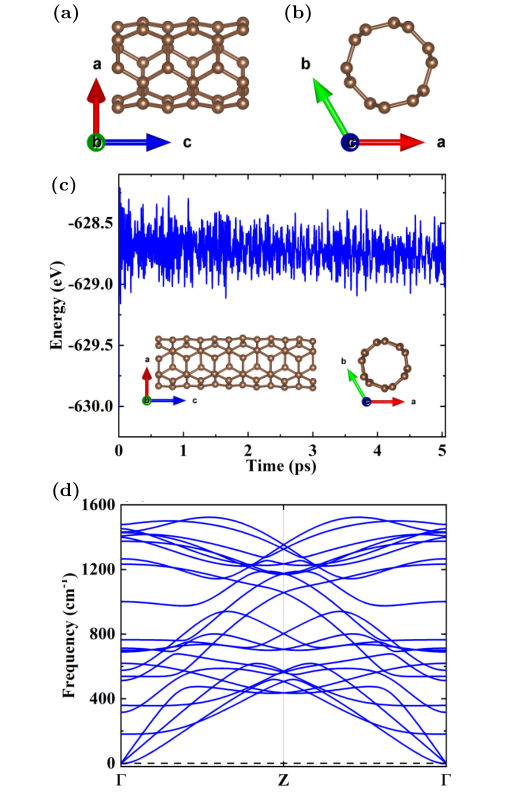}		
			\end{overpic}
		
	\vspace{-2mm}
	\caption{\label{fig:structure}
		(a) and (b), Side view along the $b$ axis and top view along the $c$ axis of the (3,3) carbon nanotube structure.
		(c) Total energy evolution during molecular dynamics simulation; insets show the side and top views of the final configurations at 300~K after 5~ps.
		(d) Phonon spectra of the (3,3) carbon nanotube.
	}	
	\end{minipage}
\end{figure}

The atomic structure of (3,3) carbon nanotube are shown in Figs.~\ref{fig:structure}(a) and \ref{fig:structure}(b), in which the C-C bond length is 1.44 \AA\ and the diameter is 4.20 \AA\ for the optimized structure.
In our calculations, a primitive cell contains twelve carbon atoms and the lattice parameters are $a$ = 22.42 \AA, $b$ = 22.42 \AA, $c$ = 2.46 \AA.
Large values of  $a$ and $b$ can ensure that the interactions between nanotubes in adjacent unit cells are negligible.

The thermal and dynamical stability of the optimized structure was examined through \textit{ab initio} molecular dynamics and phonon calculations.
As shown in Fig.~\ref{fig:structure}(c), the total energy fluctuates around a certain value at 300~K and no significant drop in energy is observed.
The insets in Fig.~\ref{fig:structure}(c) show the final configurations at 300 K after 5 ps, confirming that the basic structure of carbon nanotube is preserved and there is no structual collapse.
The phonon dispersion shown in Fig.~\ref{fig:structure}(d) exhibits no appreciable imaginary modes across the Brillouin zone.
A very small negative frequency appears only in the vicinity of the $\Gamma$ point, with a magnitude below $1.6~\mathrm{cm^{-1}}$, which is well within the numerical noise typically associated with DFPT calculations.
The absence of imaginary mode in the entire Brillouin zone indicates that (3,3) carbon nanotube is dynamically stable.

\subsection{\label{sec:level2}Phonon spectra and electron-phonon coupling of the (3,3) carbon nanotube with 4.5\% tensile strain}

\begin{figure*}[t]
	\centering
		
		\begin{overpic}[height=5.8cm,trim=0 0 0 0,clip]{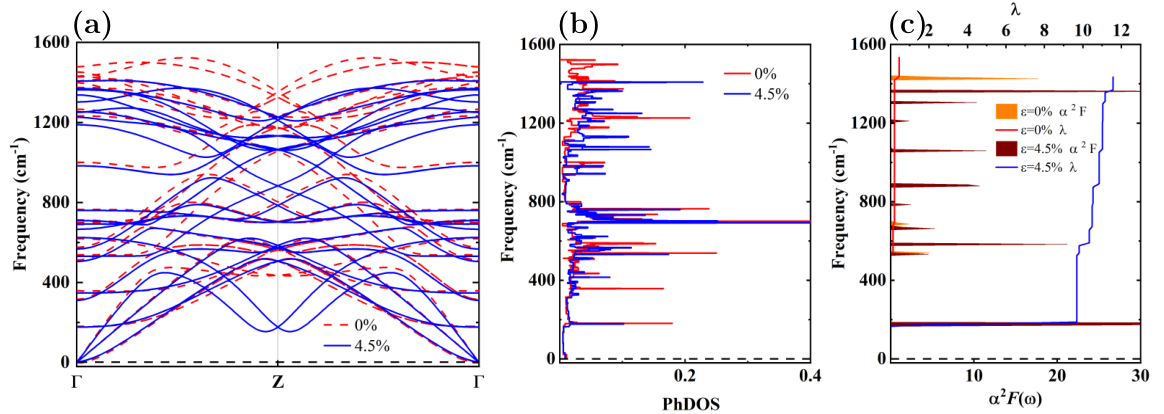}
		\end{overpic} 	
	\caption{\label{fig:pho-phdos-lambda-compare}
		Comparison of phonon and EPC properties of the (3,3) carbon nanotube in the unstrained state (red) and under 4.5\% tensile strain (blue).
		(a) Phonon dispersion.
		(b) Corresponding PhDOS.
		(c) Eliashberg spectral function $\alpha^{2}F(\omega)$ together with the cumulative electron--phonon coupling parameter $\lambda(\omega)$.
	}
\end{figure*}

The blue lines in Fig.~\ref{fig:pho-phdos-lambda-compare}(a) represent the phonon dispersion of the (3,3) carbon nanotube under 4.5\% tensile strain. No imaginary frequencies are detected, confirming the dynamic stability of the (3,3) carbon nanotube under this tensile strain condition.
For comparison, the phonon dispersion of unstrained nanotube is also presented with the red dashed lines in Fig.~\ref{fig:pho-phdos-lambda-compare}.
A systematic downward shift of the entire phonon spectrum is observed under tensile strain, with the most pronounced softening occurring in the low-frequency acoustic modes (below 800 cm$^{-1}$). This phonon softening reflects a reduction in the restoring force of atomic vibrations under tensile strain.

This phonon softening is further quantified in Fig.~\ref{fig:pho-phdos-lambda-compare}(b), which shows the corresponding phonon density of states (PhDOS). Under 4.5\% tensile strain (blue lines), the PhDOS exhibits a substantial increase in the low-frequency region (0-800 cm$^{-1}$), while the density of high-frequency modes (800-1600 cm$^{-1}$) decreases. This redistribution of phonon states directly stems from the frequency downshift observed in the dispersion curves, as high-frequency modes are softened into the low-frequency regime. The accumulation of low-frequency phonons is particularly critical for EPC enhancement.

The impact of these phonon changes on EPC strength is explicitly demonstrated in Fig.~\ref{fig:pho-phdos-lambda-compare}(c). The Eliashberg spectral function $\alpha^{2}F(\omega)$ reveals a dramatic increase in low-frequency spectral weight under tensile strain (blue curve), indicating a much stronger interaction between electrons and low-frequency phonons. Concomitantly, the cumulative electron-phonon coupling parameter $\lambda(\omega)$ rises from 0.4 in the unstrained state (red curve) to 11.3 under 4.5\% strain, confirming an order-of-magnitude enhancement in EPC strength.

\subsection{\label{sec:level2} Estimation of superconducting critical temperature}

\begin{table*}[t]
	\caption{\label{tab:Tc_strain}
		Calculated EPC constant $\lambda$, logarithmic average phonon frequency $\omega_{\log}$,
		second-moment phonon frequency $\bar{\omega}_2$ (both in K), electronic density of states at the $N(E_{F})$,
		and estimated superconducting critical temperatures $T_{c}$ obtained from the McMillan and Allen--Dynes equations
		for the (3,3) carbon nanotube under different tensile strains.}
	\begin{ruledtabular}
		\begin{tabular}{cccccccccc}
			\textrm{Strain}
			& \textrm{$\lambda$}
			& \textrm{$\omega_{\log}$ (K)}
			& \textrm{$\bar{\omega}_2$ (K)}
			& \textrm{$N(E_{F})$}
			& \textrm{$T_{c}$ (McMillan) (K)}
			& \textrm{$f_{1}$}
			& \textrm{$f_{2}$}
			& \textrm{$T_{c}$ (Allen--Dynes) (K)}
			\\
			\colrule
			0\%  & 0.49 & 1340 & 1575 & 4.98 & 15 & 1.02 & 1.00 & 15 \\
			2\%  & 0.65 & 1228 & 1288 & 5.44 & 36 & 1.03 & 1.00 & 37 \\
			4\%  & 4.60 & 540  & 865  & 6.04 & 113 & 1.37 & 1.29 & 200 \\
			4.5\%  & 16.73 & 327  & 504  & 6.23 & 84 & 2.29 & 1.50 & 287 \\
			5\%  & 12.23 & 298  & 560  & 6.42 & 74 & 1.99 & 1.73 & 254 \\
			5.5\%  & 3.09 & 381  & 871  & 6.67 & 69 & 1.23 & 1.22 & 104 \\
			6\%  & 0.79 & 1064 & 1457 & 6.92 & 48 & 1.04 & 1.01 & 51 \\
			8\%  & 0.82 & 1219 & 1453 & 8.38 & 59 & 1.04 & 1.01 & 62 \\
		\end{tabular}
	\end{ruledtabular}
\end{table*}

\begin{figure}[t]
	\centering
	\begin{minipage}{0.88\columnwidth}
			\begin{overpic}[trim=0 0 0 0, clip, width=\textwidth]{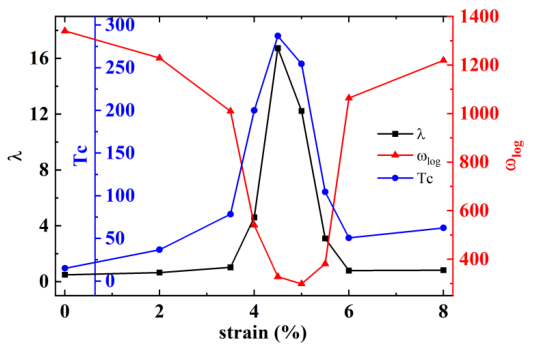}		
			\end{overpic}
		
	\vspace{-2mm}
	\caption{\label{fig:tc-strain}
		Evolution of the EPC constant $\lambda$, logarithmic phonon frequency $\omega_{\log}$, and superconducting critical temperature $T_c$ of the (3,3) carbon nanotube as functions of uniaxial tensile strain.
	}	
	\end{minipage}
\end{figure}

For the (3,3) carbon nanotube, uniaxial tensile strain can induce a non-monotonic modulation of the electron-phonon coupling strength and superconducting critical temperature.
We consider a range of uniaxial tensile strains from 0\% to 8\%, and the calculated results are summarized in Table~\ref{tab:Tc_strain} and Fig.~\ref{fig:tc-strain}. These values are derived with the computational parameter 1$\times$ 1 $\times$ 96 $k$-mesh and 1$\times$ 1 $\times$ 4 $q$-mesh.
To determine a reasonable value of the smearing parameter $\sigma$, the relevant calculations were performed, as detailed in the Appendix A.
The unstrained nanotube exhibits weak EPC, with EPC constant $\lambda = 0.49$ and a high logarithmic phonon frequency $\omega_{\log} \approx 1340$ K, which yields a relatively low critical temperature of $T_{c} \sim 15$ K.
A small tensile strain of 2\% slightly enhances both $\lambda$ and the electronic density of states at the $N(E_{F})$, producing a modest increase in $T_{c}$ to $37$ K.
At 4.5\% tensile strain, the EPC constant undergoes a dramatic enhancement, increasing to $\lambda = 16.73$.
This sharp jump is accompanied by a substantial reduction of the logarithmic phonon frequency from $\sim 1340$K to $\sim 327$K, signaling pronounced softening of specific optical and transverse modes that strongly enhance EPC.
The electronic density of states at the $E_{F}$ also increases to $N(E_{F}) = 6.23$, further amplifying the coupling strength.
 Using the McMillan formula, $T_{c}$ is calculated to be 84 K, while the Allen--Dynes formula, including strong-coupling and shape corrections ($f_{1} = 2.29$, $f_{2} = 1.50$), gives a much higher $T_{c}$ of 287 K.
At 6\% strain, the EPC decreases to $\lambda = 0.79$ and $\omega_{\log}$ recovers to above 1000K, while $T_{c}$ is reduced to 51 K.
At 8\%, the value of density of states at the $E_{F}$ increases to 8.38, but the EPC is $\lambda = 0.82$, leading to $T_{c}$ values of only 62 K.
At this stage, we have demonstrate that the $4.5\%$ tensile strain is the optimal condition for superconductivity in the $(3,3)$ carbon nanotube.

\begin{table*}[htbp]
\centering
\caption{Calculation results with different q-mesh settings}
\label{tab:qmesh_results}
\begin{ruledtabular}
\begin{tabular}{lccccccccccc}
Q-mesh & strain & $\sigma$ (Ry) & $\lambda$ & $\omega_{log}$ (K) & $\bar{\omega}_2$ (K) & $N(E_{F})$ &  $T_{c}$ (McMillan) (K) & $f_{1}$ & $f_{2}$ & $T_{c}$ (Allen-Dynes) (K) \\
\hline
1 $\times$ 1 $\times$ 4  & 4.5\%  & 0.005         & 16.73  & 327.45         & 504.10     & 6.23  & 83.64    & 2.29 & 1.50 & 287.10     \\
1 $\times$ 1 $\times$ 8  & 4.5\%  & 0.005         & 10.04  & 327.45         & 504.10     & 6.23  & 79.58    & 1.83 & 1.45 & 210.19     \\
1 $\times$ 1 $\times$ 12 & 4.5\%  & 0.005         & 7.17  & 327.88         & 514.10     & 6.23  & 75.78     & 1.60 & 1.40 & 169.27     \\
1 $\times$ 1 $\times$ 16 & 4.5\%  & 0.004         & 6.84  & 328.54         & 507.41     & 6.23  & 75.30     & 1.57 & 1.38 & 162.54     \\
\end{tabular}
\end{ruledtabular}
\end{table*}
The obtained 287 K is too high, which raises concerns about the reliability of the calculated data. We thus increased the q-mesh density to re-examine this issue.
The results obtained with a denser $q$-mesh are presented in Table~\ref{tab:qmesh_results}. When the $q$-mesh increase from 1 $\times$ 1 $\times$ 12 to 1 $\times$ 1 $\times$ 16, the EPC constant $\lambda$ decreases from 7.17 to 6.84 and the superconducting critical temperature decreases from 169.27 K to 162.54 K. Since the variation has become relatively small, and taking computational cost into account, we regard 162.54 K as a reasonable result.


\section{\label{sec:level1}Conclusion}
Based on first-principles calculations, we study the electronic structure and the electron-phonon coupling in the quasi-one-dimensional (3,3) carbon nanotube.
 Our results clearly demonstrate the significant impact of uniaxial tensile strain on the EPC and superconducting critical temperature of the (3,3) carbon nanotube.
When a moderate tensile strain of approximately 4.5\% is applied, the system undergoes a synergistic effect of overall phonon softening and increased electronic density of states at the Fermi level, promoting the electron-phonon coupling strength and pushing the superconducting critical temperature up to about 162 K.
As a result, our work not only suggests a practical and experimentally accessible pathway for tuning superconductivity in carbon nanotubes without the need for extreme pressure conditions, but also theoretically predicts that high-temperature superconductivity can occur in stretched carbon nanotubes.
	
\begin{acknowledgments}
This research was funded by the National Natural Science Foundation of China under Grants Nos. 12274458, 12274255.
\end{acknowledgments}

\appendix
\section{Convergence tests of the EPC constant}

	
	
		
%
		

The accurate evaluation of the EPC constant $\lambda$ in low-dimensional metallic systems is numerically demanding, primarily due to the presence of double Dirac delta functions in the EPC matrix elements. In practical DFPT calculations, these delta functions are approximated by smearing functions characterized by a finite broadening width $\sigma$. Consequently, the computed $\lambda$ depends sensitively on both the smearing parameter $\sigma$ and the sampling density of the Brillouin zone.

From a formal perspective, the physically meaningful EPC constant corresponds to the joint limit $\sigma \rightarrow 0$ and the number of $\mathbf{k}$ points $N_k \rightarrow \infty$.
As a result, insufficient $\mathbf{k}$-point sampling or excessively large smearing may lead to artificial suppression or enhancement of $\lambda$.

\begin{figure*}[htbp]
	\centering
	\begin{minipage}{1.8\columnwidth}
			\begin{overpic}[trim=60 60 60 68, clip, width=\textwidth]{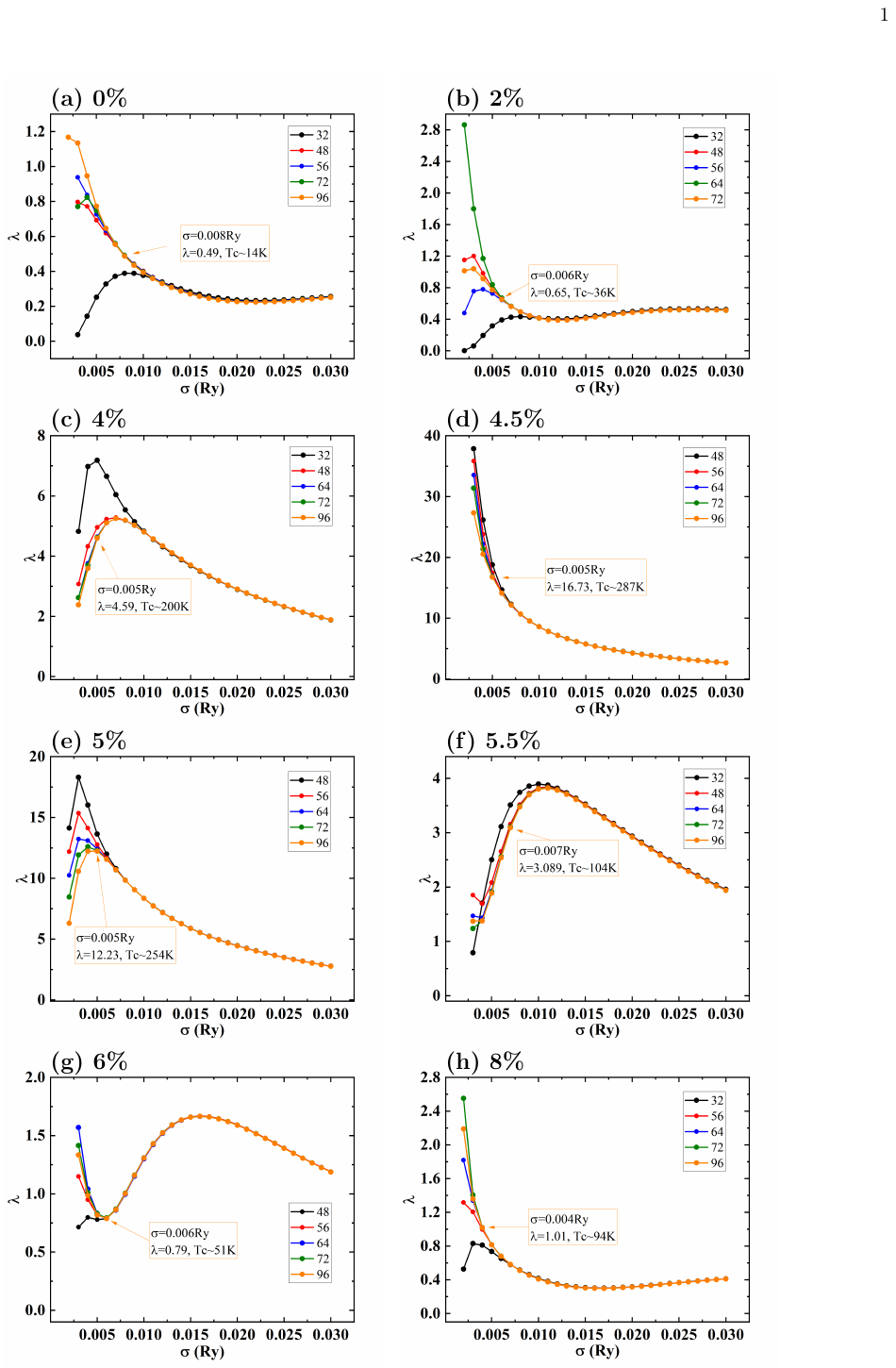}		
			\end{overpic}	

	\caption{\label{fig:EPC_convergence}
		Convergence of the EPC constant $\lambda$  as functions of the electronic smearing width $\sigma$ for the (3,3) carbon nanotube under uniaxial tensile strains of 0\%, 2\%, 4\%, 4.5\%, 5\%, 5.5\%, 6\%, and 8\%.
		The legend in each panel specifies the axial $\mathbf{k}$-point sampling $1\times 1\times N_k$ used for the Brillouin-zone integrations, while the phonon and EPC calculations employ a fixed $1\times 1\times 4$ $\mathbf{q}$-point mesh for all strains.
	}
\end{minipage}
\end{figure*}

To ensure the reliability of the EPC results for the $(3,3)$ carbon nanotube under uniaxial tensile strains from $0\%$ to $8\%$, we carried out systematic convergence tests with respect to both the smearing width $\sigma$ and the $\mathbf{k}$-point sampling density, following the established procedure adopted in previous EPC studies based on Wannier interpolation \cite{gao2017prediction}. The results are summarized in Fig.~\ref{fig:EPC_convergence}.
Several $\mathbf{k}$-point meshes along the tube axis, ranging from $1 \times 1 \times 32$ to $1 \times 1 \times 96$, were examined in combination with a series of smearing widths.

Our results demonstrate that, for all considered strain values, the EPC constant $\lambda$ exhibits pronounced sensitivity to $\sigma$ when coarse $\mathbf{k}$ meshes are employed. In contrast, upon increasing the $\mathbf{k}$-point density, $\lambda$ progressively stabilizes over a finite range of small $\sigma$. In particular, the fine meshes of $1 \times 1 \times 72$ and $1 \times 1 \times 96$ yield nearly identical $\lambda$ values for $\sigma \approx 0.005$~eV, indicating that the EPC constant is effectively converged with respect to $\mathbf{k}$ sampling in this regime.

It is worth noting that commonly adopted smearing parameters in the literature, typically in the range of $0.01$--$0.02$~Ry, may remain too large to faithfully represent the $\sigma \rightarrow 0$ limit for one-dimensional metallic systems. In contrast, the smaller smearing value $\sigma \approx 0.005$~eV employed in the present work is substantially closer to this limit, while still ensuring numerical stability of the EPC calculations.

Based on the above convergence analysis, we adopt the EPC quantities obtained using a $1 \times 1 \times 96$ $\mathbf{k}$ mesh and $\sigma \approx 0.005$~eV as the final results reported in the main text. This choice provides a reliable balance between numerical accuracy and computational feasibility, and guarantees that the reported strain-dependent EPC properties of the $(3,3)$ carbon nanotube are free from spurious smearing or sampling artifacts.

\bibliographystyle{apsrev4-2}
\bibliography{CNT_references}

\end{document}